# Case Frames and Case-Based Arguments in Statutory Interpretation

Michał Araszkiewicz[a,1]

[a]*Jagiellonian University in Kraków, Department of Legal Theory*
ORCiD ID: 0000-0003-2524-3976

**Abstract.** We introduce a novel conceptual Case Frame model that represents the content of cases involving statutory interpretation within civil law frameworks, accompanied by an associated argument scheme enriched with critical questions. By validating our approach with a modest dataset, we demonstrate its robustness and practical applicability. Our model not only provides a structured method for analyzing statutory interpretation but also highlights the distinct needs of lawyers operating under statutory law compared to those reasoning with common law precedents. The model presented here is a step towards developing a hybrid Machine Learning–Argumentation system that includes a module for constructing well-structured arguments from textual datasets

## 1. Introduction

Case-based reasoning (CBR) is essential for legal argument in jurisdictions such as the United States and England, where *stare decisis* applies, and earlier judicial decisions formally constrain subsequent ones. Consequently, Anglo-American legal practice involves identifying relevant case similarities and differences. If the current factual situation is relevantly similar to a binding precedent, the precedent should generally govern unless sufficient differences can be identified through successful distinguishing [1]. Much of AI and Law research is dedicated to understanding the reasoning behind assessing similarities and differences between cases. This is particularly evident in projects like HYPO and its progeny [2, 3] and in formal analyses of precedential constraint, such as those by [4] and subsequent works developing the reason model of precedent [5, 6] summarised in [7]. The principal approach to representing cases is through a set of factors, commonly referred to as 'stereotypical fact patterns,' initially considered to be attached to specific outcomes [8] and later, in more refined models, to issues [9].

AI and Law research has paid much less attention to using case-based argumentation to justify conclusions about statutory interpretation within the context of the civil law tradition. Instead, general models of statutory interpretation were developed and studied [10]. This approach is natural and justified because, in civil law jurisdictions, interpretive arguments based on previous cases are one type of argument among many [11]. However, their significance should not be underestimated. Although the conclusions of such arguments are generally not binding, they carry significant persuasive force, especially if the highest courts decide the cited cases. Moreover, even in civil law jurisdictions, precedents can have a degree of formal binding force based on specific regulations. In practice, therefore, it is relatively challenging to succeed in arguing directly against the statutory interpretations accepted in the highest courts' decisions. Therefore, to develop practical tools supporting lawyers (including judges) in statutory interpretation tasks, examining the structure of interpretive knowledge extracted from cases and the arguments based on this knowledge is essential.

---

[1] Corresponding Author: michal.araszkiewicz@uj.edu.pl. This research was funded by National Science Centre, Poland (grant no. 2023/49/B/HS5/00753).

Importantly, factor-based knowledge and reasoning are limited in case-based arguments for statutory interpretation rooted in civil law culture. Although assessing similarities and differences between states of affairs is present in reasoning, the applicability of particular statutory rule conditions ultimately groups cases. Even if a given factual situation is considered viable for satisfying a rule's conditions, the outcome typically depends on interpreting those conditions. The interpretation of the rule's conditions ultimately determines whether a given state of affairs satisfies the rule.

Notably, in the civil law tradition, reasoning does not typically proceed by generalizing from facts through intermediate concepts to issues [as in 12]. Instead, once the set of facts determined by evidentiary proceedings is fixed, the focus shifts to questions concerning a rule's interpretation, determining its eventual application or lack thereof. Therefore, a model for case-based argumentation in statutory reasoning should maintain a similar focus.

Two crucial questions emerge:
1) What is the structure of knowledge stored in cases essential for developing a case-based argument for an interpretive conclusion?
2) What argument structure is supported by these knowledge units?

## 2. Legal-Theoretical Background. Conceptual and Methodological Issues

*2.1. What is Interpretation?*

Much legal-theoretical work is devoted to, or arises from, problems of legal interpretation [13]. However, the very notion of legal interpretation remains debatable. A widely accepted opinion, also present in legal practice, holds that the interpretation of a statutory expression consists of ascribing meaning to that expression [10] or, more narrowly, resolving doubts concerning the ascription of meaning [14]. The understanding of these claims depends on the assumed theory of meaning.

Two opposite strategies characterize how a legal theory of interpretation may deal with doubts about the concept of meaning. Importantly, these strategies are not a dichotomy but rather extreme points on a continuum. One extreme is to adopt a robust theory of meaning developed outside jurisprudence—for example, in the philosophy of language, cognitive linguistics, or literary studies; for an instructive example of using a philosophical theory of meaning in legal interpretation theory, namely, Kripke-Putnam semantics, see [15]. Such a solution cannot serve direct descriptive functions because lawyers do not adhere to any robust theory of meaning in their communications about legal interpretation. They typically lack training in these theories, and their potential adherence to them, primarily philosophical ones, would be doubtful from the point of view of the rule of law.

The opposite strategy, here referred to as the flat one, does not attempt to define meaning or interpretation. Instead, it directs researchers to investigate the reasoning communicated by lawyers and to reconstruct these concepts based on the examined material. Such models may be more descriptively adequate but lack generality and coherence due to a lack of deeper foundations. In developing our model, we adopt a position closer to the flat strategy, focusing on knowledge stored in and extractable from judicial opinions.

For our purposes, it suffices to acknowledge that "meaning" has a dual sense: it may indicate either the set of criteria (possibly vague or contested) used to establish the set of objects to which a linguistic expression is applicable (the intensional perspective), or simply the set of such objects (the extensional perspective) (for the distinction: [16], applications of extensional perspective: [10, 17]). Both are present in arguments communicated in judicial decisions. In some cases, where relatively general predicates characterize the set of objects covered by a statutory expression, it is difficult to determine whether a given interpretive statement focuses on intension or extension. However, incorporating the distinction is vital because framing an interpretation as more intensional or extensional has significant practical consequences. Suppose an intensional interpretive statement in a landmark case provides a set of criteria for interpreting an expression. In that case, there is a strong argument for using these criteria in interpreting that expression. Conversely, if an extensional interpretive statement is formed, the criteria for classifying an object as an instance of an expression typically remain debatable.

In both cases, an interpretive statement can be understood as establishing a relation between a statutory expression (hereafter called *interpretandum*) and another linguistic expression (*interpretans*). The content of the *interpretans* will represent either a set of criteria (in the case of an intensional interpretive statement) or a set of objects or states of affairs (in the case of an extensional interpretive statement). In the latter case, the relation between *interpretandum* and *interpretans* may be understood as set-theoretical equivalence or inclusion [as in 10 and 17]. Importantly, interpretive statements do not have to be exhaustive; they may provide only exemplary classification criteria or designates of an expression. Moreover, interpretive statements may be positive or negative—a given criterion may explicitly be held inapplicable, or an object may be classified as outside the scope of an expression.

*2.2. How are Interpretive Statements Justified?*

The question above has a widely accepted answer: interpretive statements are justified if they are conclusions of well-constructed interpretive arguments [10, 11]. The stronger the argument, the higher the degree of justification. Traditionally, interpretive arguments are based on interpretive canons [10]. Let the capital letters **E**, **D**, **M**, and **C** represent certain expressions, documents, meanings, and canons, respectively. In contrast, the lowercase letters **e**, **d**, **m**, and **c** indicate specific expressions, documents, meanings, and canons. A general scheme of an interpretive argument can be reconstructed as follows [10]:

**Universal Argument Scheme for Statutory Interpretation**

**Major premise:** If the interpretation of **E** in **D** as **M** satisfies **C**'s condition, then **E** should (not) be interpreted as **M** in **D**.

**Minor premise:** The interpretation of **e** in **d** as **m** satisfies **c**'s condition.

**Conclusion (interpretive statement): e** should (not) be interpreted as **m** in **d**.

There are different taxonomies of interpretive canons; for example, one distinguishes between linguistic, systemic, functional, and "transcategorical" arguments [11]. Justifying an interpretive statement is relatively straightforward if all applicable canons support it. However, the situation becomes problematic when different canons support incompatible conclusions. Legal theory has developed tools to reduce the complexity of such situations: second-order interpretive directives. These directives govern the ordering of the application of arguments based on canons (directives of procedure) and assign default greater strength to specific arguments (directives of preference). An account of such directives can be found in the quoted source [11, p. 531-532], which advocates accepting precise results of linguistic interpretation unless there is a reason to employ the systemic canons. In the case of obtaining clear results on this level, again, only some reason may justify applying teleological arguments.

However, such an account of second-order directives is not the only possibility. Different approaches may be adopted in various jurisdictions, branches of law, and types of judicial proceedings; these accounts may also change over time. Recognizing and applying an appropriate model of second-order directives is an essential part of a continental lawyer's expertise.

For our flat model, the question concerning the relative strength of interpretive arguments may be expressed as follows: An interpretive argument is relatively strong if (1) using such an argument is not excluded by an applicable directive of procedure, and (2) this argument is either favored by default by an applicable directive of preference, or there exist reasons to assign it greater strength than other arguments, including those favored by the applicable directive of preference. The reasons that could have such an effect depend on the context of the case in question, the analyzed legal system, the branch of law to which the specific provision belongs, the goal of the regulation, the type of interpreted provision, etc. [18] The level of generality of these findings is too high to provide specific support in practical matters; hence, lawyers are naturally inclined to look for answers to interpretive questions in previously decided cases.

## 3. Case Frame and Appeal to a Prior Case Argument Scheme

As noted above, in the context of statutory law interpretation, lawyers are first and foremost interested in learning how (a particular expression of) a legal provision was interpreted before and how this interpretation was supported. The essential knowledge elements to be reconstructed from previous cases are, therefore, as follows [19].

**Def 1. A Case Frame for Statutory Interpretation** is a four-tuple consisting of:

**Part 1. Case Data.** Characterization of a case in which the knowledge is extracted from, which is a five-tuple:

<*Jurisdiction, Court, Identifier, Date, Procedural*>, where *Jurisdiction* is a slot assuming a value from a range of jurisdictions (countries and, where appropriate, also lower geographical units), *Court* is a slot representing the name of a court which enacted a decision, *Identifier* represents a formal identifier of the case according to a convention adopted in a given jurisdiction, *Date* means a date of the decision and *Procedural* gives an information concerning the status of the case, in particular whether it is a final decision.

**Part 2. Winning Interpretation.** This part of the knowledge model encompasses information about the interpretive statement adopted by the case in question. It is a seven-tuple consisting of the following elements:

<*Document, Characteristics, Interpretandum, StateOfAffairs, Interpretans, InterpretansType, Canon*>, *Document* represents the identifying data of the source of law containing *Interpretandum*, *Characteristics* represents the features of this source of law which may have a bearing on the process of interpretation, *Interpretandum* represents the interpreted (possibly: complex) expression with indication of its systematic unit in the source's text, *StateOfAffairs* is a set of formulas representing established facts of the case, *Interpretans* is a slot to be filled by the phrase representing meaning (in intensional or extensional sense) ascribed to *Interpretandum*, *InterpretansType* assumes one of two values: intensional or extensional, and *Canon* represents a set (possibly containing one element) of canons supporting the interpretation adopted by the court. Let us note that sometimes courts characterize canons very generally.

**Part 3. Defeated Interpretations.** This part represents interpretations rejected by the court. Its structure is analogous to Part 2, but as this part concerns different interpretations of the same expression, it is unnecessary to repeat the first three slots. Note also that the *Interpretans* slot may contain more than one element here.

**Part 4. Second-order Directive and its Context.** This part consists of three elements:

<*SecondOrderDirective, Context*>, representing, first, a second-order directive (either a directive of procedure, a directive of preference, or both) used by the court to resolve any conflicts between competing interpretive arguments, and second, contextual information appropriate for application of this second order directive in a way that favored Winning Interpretation; for example, if a second-order directive authorizes departure from the linguistic interpretation and favoring a teleological one only for "important reasons", the slot will indicate what were the crucial reasons recognized by the court.

Let us present an example of a filled Case Frame based on an actual judicial decision.

**Table 1.** An example of a Case Frame.

| Part 1. Case Data | | |
|---|---|---|
| | Jurisdiction | Poland |
| | Court | Supreme Administrative Court |
| | Identifier | II FSK 2051/10 |
| | Date | 21 April 2011 |
| | Procedural | Final |
| Part 2. Winning Interpretation | | |
| | Document | Regulation of the Council of Ministers of 14 September 2004 (Journal of Laws No. 218, item 2209) |
| | Characteristics | Tax law, income tax exemption, goal: improvement of the economic situation in the region |
| | Interpretandum | Expression "incurring the cost", par. 4 of the Regulation |
| | StateOfAffairs | Company documented the cost and intends to apply for tax exemption |
| | Interpretans | Documenting and recording the cost in the company's books |
| | InterpretansType | Extensional |
| | Canon | Systemic, historical, teleological |
| Part 3. Defeated Interpretations | | |
| | Interpretans | Incurring actual cost |
| | InterpretansType | Extensional |
| | Canon | Linguistic |
| Part 4. Second-order Directive and its Context | | |
| | Second-order Directive | When interpreting the law, the interpreter must not completely ignore the systemic or functional interpretation by limiting himself solely to the linguistic interpretation of a single provision. |
| | Context | Coherence with accounting regulation |

The disputed interpretive issue concerned whether the expression "incurring the cost" should be understood as "the actual incurrence of the cost" (this was the position of the authority) or as "documenting and recording the cost in the company's books" (this was the position of the taxpayer, eventually accepted by the Supreme Administrative Court). The SAC ruled for the taxpayer, adopting a holistic second-order interpretive rule, particularly seeking harmony with understanding the term "incurring the cost" in the interpreted Regulation and the accounting provisions. In this respect, this holding may be found surprising, as textbook knowledge teaches that in the field of tax law, especially regarding tax exemption provisions which have the status of exceptions from a general rule, strict linguistic interpretation should be preferred. Moreover, the SAC underlined that although linguistic canons somewhat support competing interpretations, the winning one finds additional support from different canons.

Let us comment briefly on the essential character of knowledge stored in the Case Frame. Part 1 consists, at first glance, of metadata only, but as we will see below, all these data, except for the *Identifier*, may be effectively used in critical argumentation. The first four slots of Part 2 – that is, *Document, Characteristics, Interpretandum* and *StateOfAffairs* are crucial in connection with establishing and evaluation of similarity relation between the cited case and the current fact situation (the problem). Importantly, these features of the problem will presuppose a solution to interpretive problems in *p* and, therefore, have to be determined in advance, at least tentatively [14]. The *Interpretans* and *InterpretansType* slots give crucial information about an interpretive conclusion in c, and *Canon* – on the arguments supporting it. We also obtain analogous information about defeated arguments. Finally, *Second-order Directive* slot presents justification for such a result of "argument battle". A lawyer attempting at solving p, may venture to transfer any of the known values of slots in Case Frame for c to the empty slots in Case Frame for p, provided that the argument to this effect will be well-structured.

The knowledge stored in a case frame may be used to reconstruct a specific argumentation scheme for statutory interpretation based on a cited case. We use small letters to indicate that the slots are assumed to be filled already.

**Appeal to a Prior Case Argument Scheme**

**Premise 1.** In a case *c*, characterized by Case Data *<jurisdiction, court, name, date, procedural>*, it was held that according to a Winning Interpretation, *interpretandum,* in *document* having features *characteristics*, should be ascribed with *interpretans* having the *interpretanstype* on account of *canon*, where *secondorderdirective* favours Winning Interpretation in the *context*.

**Premise 2 (Similarity).** The Case Frame for the current fact situation, *p*, shares at least one of the elements, *α*, present in *document, characteristics, interpretandum* or *stateofaffairs* slots of the Case Frame for *c*.

**Conclusion.** Another element from the Case Frame for *c*, *β*, should (not) be included in the Case Frame for *p*.

This argument scheme specifically represents a canon based on an earlier judicial decision. Still, it is also generalized to cover different uses of references to past cases in actual practice. A reference to an earlier case (here represented by a Case Frame) may be used not only to argue that a specific interpretive statement should be adopted because of some similarity feature between the cases. It may also be used to transfer any other elements (except those already fixed in the current fact situation Case Frame) since there is another similarity feature between cases. For example, on account that certain *interpretans* of an *interpretandum* was held to be supported by linguistic canon in the cited case *c*, it may serve as a basis for the contending that certain *interpretans* of an *interpretandum* should also be held to be supported by a linguistic canon in current fact situation, *p* - for instance, because in both cases, a criminal law provision is interpreted. Or, if in the current fact situation *p* it is not clear what second-order directive should be applied, a second-order directive may be taken from the Case Frame for the cited case, *c*. In sum, the above argument scheme is a concretization of a general argument scheme from analogy, as discussed in logical textbooks [20].

Of course, such an argument may be attacked on different grounds. Let us enumerate typical critical questions lawyers use in connection with such arguments (cf. [21]).

CQ1. (Similarity relevance). Is a feature f, shared by c and p, actually relevant?

CQ2. (Distinguishing). What are the differences between *c* and *p*?

CQ2a. (Branch of law). Does the *interpretandum* in *c* belong to the same branch of law as the *interpretandum* in *p*?

CQ2b. (Provision type). Is the *interpretandum* in *c* a part of the provision of the same type as the provision the *interpretandum* in *c* is a part of?

CQ2c. (Goal). Is the *interpretandum* in *c*, and the provision and document it is contained in, directed to reach the same goal as the *interpretandum* in *p*, and the provision and document it is contained in?

CQ3. (Counterexample). Is there a case *r* which shares more common features with *p* than the cited case *c*, and the Case Frame for *r* does not contain the element β?

CQ4. (Jurisdiction). Considering the jurisdiction case where *c* was enacted, should *c* have any relevance for reasoning in *p*?

CQ5a. (Obsoleteness). Considering a relatively distant date on which case *c* was enacted, should c have any relevance for reasoning in *p*?

CQ5b. (Recency). Considering a relatively recent date on which case *c* was enacted, should *c* be considered so well-established to influence reasoning in *p*?

CQ6. (Court hierarchy). Was *c* decided by a court relatively higher in the hierarchy than the court to decide the case *p*?

CQ7. (Procedural considerations). Are there any procedural considerations concerning *c*, for instance, its non-finality, that could affect its influence in *p*?

CQ8. (Other second-order rules). Are there any cases that share a relevant common feature with *p* that use a different second-order rule than *c*?

Let us add that constructing such arguments in a case base may have an iterative nature. Consider that a case *m* exists in the case base and that a case *n* was decided in a particular manner because of similarity to *m*. Now, in the Case Frame for *n,* a slot exists where canon "appeal to a prior case" is indicated as supporting the Winning Interpretation. This enables *nested* references to prior cases: in a new case *o*, similar to *n*, it is now possible to refer to a case that used an argument based on a prior similar case *n*, and on the case *n* referred to (that is, *m*). In such a way, the so-called stable lines of judicial opinions are created.

## 4. Validation by a Dataset

To validate the robustness of the proposed approach, a dataset comprising ten randomly selected decisions enacted by the Supreme Administrative Court of Poland invoking the term "linguistic interpretation" was created. The set of cases was manually annotated using the type system following the list of elements presented in the Case Frame. The partial results concerning 5 cases (due to space limitation), representing case identifiers, the canon supporting the Winning Interpretation and the applied Second-order directive, are presented in the following table.

**Table 2.** Selected Case Frames' elements identified in the dataset.

| No. | Case identifier | Canon supporting the Winning Interpretation | Second-order directive |
|---|---|---|---|
| 1. | I OSK 1714/10 | Linguistic | It is only possible to depart from the clear and unambiguous literal wording of the provision and rely on other types of interpretation in exceptional situations, when there are particularly important reasons for doing so. |
| 2. | II GSK 2177/11 | Linguistic and systemic | In the first place, apart from linguistic interpretation, a systemic interpretation should be applied. Only then, if it turned out to be impossible to interpret the concept using linguistic and systemic interpretation methods, it would be justified to refer to concepts from outside (the branch of law) |
| 3. | II OSK 725/06 | Linguistic and teleological | It is not permissible to apply a linguistic interpretation in isolation from a purposive and functional interpretation. |
| 4. | II FSK 2801/13 | Linguistic | Systemic interpretation is considered subsidiary or supporting - it is used to resolve doubts raised by linguistic interpretation and only in exceptional situations is it the basis for correcting the result of linguistic interpretation. Purpose-based interpretation is also subsidiary in nature about other interpretations - linguistic and systemic. |
| 5. | I OSK 3106/12 | Linguistic | An exceptional legal regulation cannot be subject to extensive interpretation by departing from the rules of linguistic interpretation. |

Even these partial results reveal a significant diversity regarding the formulation of the Second-order Directive. Let us note that although in case 5, its formulation has a limited range (applicable only to specific legal regulations, creating exceptions), the four other formulations have more universal ambitions and appear incompatible. A case base encompassing only those 5 cases could provide a basis for contrary interpretative arguments in any current fact situation, sharing at least one relevant similarity feature with each of the cases stored in the case base.

**5. Discussion and Related Work**

This paper provides a conceptual basis for formalizing legal knowledge and reasoning relevant to statutory interpretation. By employing the theory of argumentation schemes—including argument schemes and critical questions—the reasoning patterns presented here can be formalized in any well-founded structured argumentation system [22, 23, 24]. A more challenging task is the formal representation of relevant knowledge, presented here as Case Frames. While it is possible to apply the ANGELIC II methodology [25] to represent domain knowledge and describe the knowledge units stored in Case Frames, it would be necessary to adapt the method to encompass these units appropriately. Moreover, it is worth noting that second-order interpretive directives seem to have a cross-domain character—some are deemed universally applicable—and within a single domain, several incompatible second-order interpretive directives may be present. Consequently, the resulting knowledge bases would exhibit local inconsistencies.

Perhaps the most significant difference between formalizations of case-based domains in common law and civil law is that classical factor-based knowledge plays a less critical role in the latter than in the former. As argued above, the essential elements of knowledge relevant to statutory interpretation are the *interpretanda* and *interpretantia*, together with applied canons and preference relations following from the second-order directives. The argument that a state of affairs is an instance of the *interpretandum* will not require many reasoning steps and thus will not necessitate extensive references to prior cases. Consequently, the reasoning typically concerns defeasible rules and preferences rather than extensive prior case references. Factor-based reasoning may play a role in interpreting particularly open-textured and context-sensitive concepts (such as "reasonable," "appropriate," etc.) [26] and in the selection of competing second-order rules [18]. Moreover, prior cases do not formally constrain subsequent decisions except specific regulations. However—as argued above—they provide an indispensable source for strong interpretive arguments and essential knowledge elements for statutory interpretation.

These contentions have significant consequences. First, unlike in the common law context, there is no general prohibition against increasing inconsistencies in the case base [27]; any lawyer or court may argue differently from earlier cases if they can find suitable arguments. Second, except for specific regulations concerning binding cases, civil law has no actual precedential constraints; thus, it is difficult to speak of a strict constraint even when two cases involve the same *interpretandum* and similar circumstances. Consequently, the interesting results concerning factor hierarchies [26, 28, 29, 30] will have limited applicability, although they may be investigated in interpreting open-textured, contextually sensitive statutory predicates.

**6. Conclusions and Future Work**

In this paper, we demonstrated how detailed knowledge of statutory interpretation from cases can be represented using a Case Frame and used to construct arguments based on prior cases. We reconstructed a detailed argument scheme with assigned critical questions. To validate the robustness of our framework, we developed and annotated a modest dataset, which confirmed that our approach aids in understanding essential elements of prior cases relevant to disputed issues of statutory interpretation.

The following steps involve formalizing the identified knowledge—starting with the annotated dataset—within a framework adapted from the ANGELIC II methodology [25], and analyzing the structure of references to prior cases within this framework. We will also test using knowledge graphs to represent dependencies between concepts in Case Frames. Furthermore, we intend to apply Natural Language Processing techniques to automatically detect Case Frame elements in textual documents, following the approaches in [31, 32]. Ultimately, we aim to implement this model in a hybrid Machine Learning–Argumentation system [33]. This would assist practicing lawyers in civil law jurisdictions who manually identify prior cases' essential elements.